\documentclass{iau}
\usepackage{graphicx}

\title[Making a Be star disk] 
{Making a Be star:\\
the role of rotation and pulsations}

\author[Coralie Neiner \& St\'ephane Mathis]   
{Coralie Neiner$^1$
\and St\'ephane Mathis$^{2,1}$}

\affiliation{$^1$LESIA, UMR 8109 du CNRS, Observatoire de Paris, UPMC, Univ. Paris Diderot\\
5 place Jules Janssen, 92195 Meudon Cedex, France\\ email: {\tt coralie.neiner@obspm.fr} \\[\affilskip]
Laboratoire AIM Paris-Saclay, CEA/DSM-CNRS-Universit\'e Paris Diderot\\
IRFU/SAp, Centre de Saclay, 91191 Gif-sur-Yvette Cedex, France$^2$\\email: {\tt stephane.mathis@cea.fr}}

\pubyear{2013}
\volume{301}  
\pagerange{1--2}
\jname{Precision Asteroseismology}
\editors{J.A. Guzik, W.J. Chaplin, G. Handler \& A. Pigulski, eds.}
\begin{document}

\maketitle

\begin{abstract}
The Be phenomenon, i.e. the ejection of matter from Be stars into a
circumstellar disk, has been a long lasting mystery. In the last few years, the
CoRoT satellite brought clear evidence that Be outbursts are directly
correlated to pulsations and rapid rotation. In particular the stochastic
excitation of gravito-inertial modes, such as those detected by CoRoT in the hot
Be star HD\,51452, is enhanced thanks to rapid rotation. These waves increase the
transport of angular momentum and help to bring the already rapid stellar
rotation to its critical value at the surface, allowing the star to eject
material. Below we summarize the recent observational and theoretical findings 
and describe the new picture of the Be phenomenon which arose from these
results.
\end{abstract}

\firstsection 
\section{Stochastic excitation of pulsations in Be stars}

Be stars are massive stars with a decretion disk that are known to pulsate
thanks to the $\kappa$ mechanism. The correlation between pulsations and the
ejections of matter into the circumstellar disk was first proposed by
\cite[Rivinius et al. (2001)]{rivinius2001} and firmly established by \cite[Huat
et al. (2009)]{huat2009} thanks to CoRoT observations.

Sub-inertial gravito-inertial modes (below twice the rotation frequency) have
recently been detected in the early Be star HD\,51452 with CoRoT \cite[(Neiner
et al. 2012)]{neiner2012}. These modes cannot be excited by the $\kappa$
mechanism usually invoked for those stars, because HD\,51452 is too hot (B0\,IVe)
to be in the $\kappa$-driven g-mode instability strip. Since the observed modes
have very low frequency and a short lifetime, we have proposed that they are
excited stochastically in the convective core and at its interface with the
surrounding radiative envelope.

In addition, low-frequency g modes have been observed with CoRoT in another
early Be star, HD\,49330, during an outburst \cite[(Huat et al.
2009)]{huat2009}. We propose that these modes are also stochastically excited,
as suggested by their short lifetime. Indeed, these modes are only visible
during the outburst, while the $\kappa$-driven p modes get destabilized during
the outburst. However, in this case, the stochastic modes we observed are
probably those excited just below the surface during the outburst rather than
the ones excited in the convective core. 

It was not expected that stochastically excited gravito-inertial modes could be
observed in massive stars \cite[(Samadi et al. 2010)]{samadi2010}. However, Be
stars are very rapid rotators and stochastic excitation is enhanced in the
presence of rapid rotation, through the Coriolis acceleration which modifies
gravity waves. This has been demonstrated analytically by \cite[Mathis et al.
(2013)]{mathis2013} and observed in numerical simulations by \cite[Rogers et al.
(2013)]{rogers2013} (see also \cite[Browning et al. 2004]{browning2004}).
Indeed, in the convective zones, when rotation is rapid, gravity modes become
less evanescent in the super-inertial regime and propagative inertial modes in
the sub-inertial regime. 

Such stochastic modes are thus probably present in all rapidly rotating massive
stars. Therefore, in the case of rapid rotators, the identification of low-frequency modes should be considered carefully and not systematically attributed
to the $\kappa$ mechanism as has been done until recently.

\section{Transport of angular momentum from the core to the surface}

\cite[Lee (2013)]{lee2013} showed that gravito-inertial modes excited by the
$\kappa$ mechanism transport angular momentum and could play a role in the Be
phenomenon. However, in the sub-inertial regime, the transport of angular
momentum was believed to become less efficient because of
gravito-inertial waves equatorial trapping \cite[(Mathis et al. 2008, Mathis
2009)]{mathis2008,mathis2009}. Our recent work shows that transport by trapped
sub-inertial waves may be sustained in rapidly rotating stars thanks to the
stronger stochastic excitation by turbulent convective flows. Moreover,
sub-inertial gravito-inertial modes have very low frequencies and therefore they
transport more angular momentum than modes with higher frequencies.

We thus propose that this mechanism allows to transport angular momentum from
the convective core of Be stars, where sub-inertial gravito-inertial modes are
excited, to their surface. The accumulation of angular momentum just below the
surface of Be stars increases the surface velocity. The surface then reaches
the critical velocity so that material gets ejected from the star.

\section{Conclusions}

Thanks to the discovery of stochastically excited gravito-inertial modes in the
hot Be star HD\,51452 and to the observation of the correlation between these
pulsations and a Be outburst in HD\,49330, both with the CoRoT satellite, we
have shown that stochastic gravito-inertial waves play an important role in Be
stars. We demonstrated analytically that it is rapid rotation that enhances
those modes in Be stars. This is also confirmed in numerical simulations. Since
sub-inertial gravito-inertial modes efficiently transport angular momentum, we
propose that they could be the key to the Be phenomenon, i.e. to the ejection of
material from the surface of Be stars into a circumstellar Keplerian disk.

\end{document}